\documentclass[12pt,reqno]{amsart}
\usepackage{indentfirst, amssymb,amsmath,amsthm,hyperref}
\usepackage{physics,breqn}
\usepackage[mathscr]{euscript}

\usepackage[caption=false]{subfig}
\usepackage{graphicx}
\usepackage{tabulary}
\usepackage{epstopdf}

\newtheorem{theo}{Theorem}

\newtheorem{rem}{Remark}

\newtheorem{cor}{Corollary}
\newtheorem{defi}{Definition}

\newcommand{\be}{\begin{equation}}
	\newcommand{\ee}{\end{equation}}
\newcommand{\beas}{\begin{eqnarray*}}
	\newcommand{\eeas}{\end{eqnarray*}}
\newcommand{\bea}{\begin{eqnarray}}
	\newcommand{\eea}{\end{eqnarray}}

\numberwithin{equation}{section}

\begin{document}
	
\setlength{\unitlength}{1mm}\baselineskip .45cm
\setcounter{page}{1}
\pagenumbering{arabic}	
\title[]{Pseudo generalized Ricci-recurrent spacetimes with certain applications to modified gravity }
	
\author[  ]{Uday Chand De$^{1}$ and Krishnendu De *$^{\,2}$  }

\address
{$^1$ Department of Pure Mathematics, University of Calcutta, West Bengal, India. ORCID iD: https://orcid.org/0000-0002-8990-4609}
\email {uc$_{-}$de@yahoo.com, ucde1950@gmail.com}
\address
{$^2$ Department of Mathematics,\\
 Kabi Sukanta Mahavidyalaya,
The University of Burdwan.\\
Bhadreswar, P.O.-Angus, Hooghly,\\
Pin 712221, West Bengal, India.\\ ORCID iD: https://orcid.org/0000-0001-6520-4520}
\email{krishnendu.de@outlook.in, krishnendu.de67@gmail.com }

\begin{abstract}
In this article we introduce and characterize a pseudo generalized Ricci-recurrent spacetimes. At first, we produce an example to justify the existence of such a spacetime. Then, it is provided that a pseudo generalized Ricci-recurrent generalized Robertson-Walker spacetime represents a  perfect fluid spacetime and a pseudo generalized Ricci-recurrent perfect fluid spacetime represents either a dark energy epoch of the Universe or, the velocity vector field is parallel, conservative, acceleration-free, vorticity-free, and shear-free and becomes a static spacetime. Lastly, we study the impact of this spacetime under $f(\mathcal{R})$ gravity scenario and deduce several energy conditions.
\end{abstract}
:	
\keywords{Pseudo generalized Ricci-recurrent spacetime; generalized Robertson-Walker spacetime; perfect fluid spacetime; static spacetime; $f(\mathcal{R})$ gravity.\\
*Corresponding author.}

\maketitle

\section{Introduction}

Let $M^{n}$ ($n\ge4$) be a semi (or, pseudo)-Riemannian manifold and $g$ be its semi-Riemannian metric with signature $(p,m)$, such that $p+m=n$. $M^{n}$ equipped with $g$ is referred to as a Lorentzian manifold \cite{Neil} if $g$ is a Lorentzian metric with signature $(n-1, 1)$ or, $(1, n-1)$. Time-oriented Lorentzian manifolds are Lorentzian manifolds that accept a globally time-like vector field, also physically referred to as spacetime. A generalized Robertson-Walker (shortly, GRW) spacetime is a Lorentzian manifold of dimension $n$ $\left(n\geq4\right)$ which can be shaped by a warped product $-I\times_{\varPsi^{2}}\stackrel{\ast}{M}$, $\stackrel{\ast}{M}$ stands for $\left(n-1\right)$-dimensional Riemannian manifold, $I \in \mathbb{R}$ (set of real numbers) is an open interval, and $\varPsi>0$ is the warping function. This notion was presented by Al\'ias et al. \cite{alias1} in $1995$. In particular, GRW spacetime becomes Robertson-Walker ( shortly, RW) spacetime if we assume that $\stackrel{\ast}{M}$ is a three dimensional Riemannian manifold of constant sectional curvature. For more details about GRW spacetimes, we refer (\cite{alias1}-\cite{S98}).\par

A $4$-dimensional Lorentzian manifold $\mathrm{M}^{4}$ is described as a perfect fluid spacetime (shortly, PFS) if the Ricci tensor $\mathcal{R}_{lk}$ fulfills
\begin{equation}\label{1.1}
\mathcal{R}_{lk}=\alpha g_{lk}+\beta u_{l}u_{k},
\end{equation}
$\alpha$, $\beta$ being scalars (not simultaneously zero) and the velocity vector $u_{k}$ stands for a unit time-like vector \{that is, $u_{k}u^{k}=-1$, $u^{k}=g^{lk}u_{l}$\}. The matter field in general relativity (in short, GR) is demonstrated by $T_{lk}$, which is called the energy momentum tensor (in short, EMT). Since heat conduction terms and stress terms corresponding to viscosity does not occur, the fluid is referred to as perfect \cite{HE73}. The EMT \cite{Neil} for a PFS has the following form
\begin{equation}\label{1.2}
T_{lk}=\left(\sigma+p\right)u_{l}u_{k}+p g_{lk},
\end{equation}
$p$, $\sigma$ being the isotropic pressure and the energy density, respectively. If $\sigma=p$, then the PFS is named stiff matter fluid. If $\sigma+p=0$, $p=0$ and $\sigma=3p$, then the PFS is called the dark energy epoch of the Universe, dust matter fluid and radiation era \cite{chav}, respectively. Without the cosmological constant, the Einstein's field equations (shortly, EFEs) are stated as
\begin{equation}\label{1.3}
\kappa T_{lk}-\mathcal{R}_{lk}+\dfrac{\mathcal{R}}{2}\,g_{lk}=0
\end{equation}
in which $\kappa$ and $\mathcal{R}$ stand for the gravitational constant and Ricci scalar, respectively. 
 Equations \eqref{1.1}, \eqref{1.2} and \eqref{1.3} infer that
\begin{equation}\label{1.4}
	\alpha=\kappa\left(\dfrac{\sigma-p}{2}\right)\quad\mathrm{and}\quad\beta=\kappa\left(\sigma+p\right).
\end{equation}

In $\mathrm{M}^{4}$, the Weyl conformal curvature tensor $\mathcal{C}^{l}_{ijk}$ is demonstrated by
\begin{equation}\label{1.5}
\mathcal{C}^{l}_{ijk}=\mathcal{R}^{l}_{ijk}-\dfrac{1}{2}\left\{g_{ij}\mathcal{R}^{l}_{k}
-g_{ik}\mathcal{R}^{l}_{j}+\delta^{l}_{k}\mathcal{R}_{ij}-\delta^{l}_{j}\mathcal{R}_{ik}\right\}
+\dfrac{\mathcal{R}}{6}\left\{\delta^{l}_{k}g_{ij}-\delta^{l}_{j}g_{ik}\right\}
\end{equation}
in which $\mathcal{R}^{l}_{ijk}$ denotes the curvature tensor and $\mathcal{R}^{l}_{k}=\mathcal{R}_{jk}g^{lj}$.\\

It is well-known that \cite{E49}
\begin{equation}\label{1.6}
	\nabla_{l}\mathcal{C}^{l}_{ijk}=\dfrac{1}{2}\left\{\left(\nabla_{k}\mathcal{R}_{ij}-\nabla_{j}\mathcal{R}_{ik}\right)
-\dfrac{1}{6}\left(g_{ij}\nabla_{k}\mathcal{R}-g_{ik}\nabla_{j}\mathcal{R}\right)\right\}.
\end{equation}

It is important to note that the Weyl geometry initially presented as a unifying theory of gravity and electromagnetic \cite{hj} that could challenge Einstein's theory of gravity which is the main motivation behind its investigation in this article.\par

For the purpose of determining Einstein’s field equations, the study of spacetime symmetries is important. Symmetries are characteristic of geometry and reveal physics. There are many symmetries in matter and spacetime geometry. The metric equations are useful because they make many solutions easier. In GR, their primary utilization is that they classify solutions of Einstein’s field equations.\par

In 1952, Patterson \cite{P52} invented the idea of Ricci-recurrent manifolds. In 1995, De et al. \cite{DGK95} introduced the concept of generalized Ricci-recurrent manifolds. In a recent paper \cite{mdd} Mallick et al. studied generalized Ricci-recurrent manifolds with applications to relativity. The significance of the Generalized Ricci-recurrent structure and its interaction with the modified $f(\mathcal{R})$-theory \cite{ade1} and the modified Gauss-Bonnet $f(\mathcal{R},G)$-theory \cite{ade2} are well established. On the other hand, pseudo $Z$-symmetric spacetimes have been investigated by Mantica and Suh \cite{ManSuh} and Ozen\cite{ozen} have studied $m$-projectively flat spacetimes. Also, in \cite{fku} $\Psi$-conformally symmetric spacetimes have been investigated. Moreover, in \cite{kdeu1}, we have studied $\psi$-conharmonically symmetric spacetimes. As well, many authors have looked at the spacetime of general relativity in various methods; for additional information, see ( \cite{gul2}, \cite{kdez}). Inspired by foregoing studies we introduce and characterize a novel spacetime named pseudo generalized Ricci-recurrent spacetimes.\par

If the covariant derivative of $\mathcal{R}_{lk}$ is in the following form:
\begin{equation}\label{1.7}
\nabla_{i}\mathcal{R}_{lk}=u_{i}\mathcal{R}_{lk}+v_{i}g_{lk}+w_{i}\mathcal{D}_{lk},
\end{equation}
where $u_{i}$, $v_{i}$ and $w_{i}$ are non-zero covariant vectors, $\mathcal{D}$ is the $\left(0,2\right)$-type structure tensor of the manifold obeying
\begin{equation}\label{1.8}
\mathcal{D}_{lk}=\mathcal{D}_{kl},\quad g^{lk}\mathcal{D}_{lk}=0\quad\mathrm{and}\quad\mathcal{D}_{lk}u^{k}=0,
\end{equation}
then we named such an $n$-dimensional manifold as pseudo generalized Ricci-recurrent manifold, denoted by $P\left(GR_{n}\right)$. If $w_{i}=0$, then $P\left(GR_{n}\right)$ reduces to a generalized Ricci-recurrent manifold $\left(GR_{n}\right)$ \cite{DGK95}. For $v_{i}=w_{i}=0$, $P\left(GR_{4}\right)$ becomes a Ricci-recurrent manifold $\left(R_{n}\right)$ \cite{P52}. If $\mathcal{R}_{lk}$ obeys \eqref{1.7} and $u_{i}$ is a unit time-like vector, then $\mathrm{M}^{4}$ is called a $P\left(GR_{4}\right)$ spacetime. \par
Multiplying \eqref{1.7} with $g^{lk}$ and using \eqref{1.8}, we reach
\begin{equation}\label{1.9}
\nabla_{i}\mathcal{R}=\mathcal{R}u_{i}+4v_{i}.
\end{equation}

The physical motivation for researching various spacetime models in general relativity and cosmology is to gain additional insight into particular phases of the universe's evolution, which can be divided into the following three stages:\\
(i) The initial stage, (ii) The intermediate stage and (iii) The final stage.\par
The initial stage concerns viscous fluid whereas the intermediate stage concerns non-viscous fluid and both are admitting heat flux. The final stage equipped with thermal equilibrium tells about the perfect fluid stage.
In our current study, we select the final stage.\par

The scientific world as a whole accepts the idea that our Cosmos is currently going through an accelerated phase.
EFEs are not adequate to determine the late-time inflation of the cosmos without supposing the existence of certain unseen components that could account for the dark energy and dark matter origins. It is the primary source of inspiration for the extension to acquire higher order field equations of gravity.

According to GR theory, it is commonly accepted that energy conditions (ECs) are essential resources for studying black holes and wormholes in various modified gravities (\cite{BBHL07}
 -\cite{LPC15}). The Raychaudhuri equations \cite{RBB92}, which methodically produce the ECs, express the intriguing nature of gravity through the positivity condition $\mathcal{R}_{lk}u^{l}u^{k}\geq0$, $u^{l}$ is a null vector. The last geometric criterion is the same as the null energy condition (NEC) $T_{lk}u^{l}u^{k}\geq0$. Certainly, the weak energy condition (WEC) reflects that  $T_{lk}u^{l}u^{k}\geq0,$ for every time-like vector $u^{l}$ and preserves a positive local energy density. In addition, a spacetime fulfills the dominant energy condition (DEC) if $T_{lk}u^{l}v^{k}\geq0$ holds for every two co-oriented time-like vectors $u$ and $v$ and strong energy condition (SEC) \cite{DS99} if $R_{lk}u^{l}u^{k}\geq0$ holds for all time-like vectors $u$. An overview of the four primary ECs is given here. A number of additional, lesser-known point wise ECs exist that express different restrictions on the stress energy tensor. A trace energy condition (TEC) was also present in the past. This implies that for a perfect fluid, $\sigma-3p\geq 0$. For a number of years in the 1950s and 1960s, this was thought to be a physically feasible state. Now, opinions have changed. Particularly, the TEC is violated with the finding of stiff equations of state for matter found in neutron stars \cite{tw}. We bring it up here as a specific illustration of an EC.\par

Interestingly, the idea of $f(\mathcal{R})$-gravity appears as a spontaneous extension of Einstein's theory of gravity. Here, the function $f(\mathcal{R})$, in which $\mathcal{R}$ denotes the Ricci scalar, modifies the Hilbert-Einstein action term. The aforementioned theory was invented by Buchdahl \cite{hab}, and Starobinsky \cite{aas} has demonstrated its validity through research on cosmic inflation.
Through the introduction of certain couplings between the geometrical quantities and the matter sector, the $f(\mathcal{R})$ theory of gravity has been further generalized. The non-minimal coupling between the matter lagrangian density and the curvature invariant has been proved in \cite{har}, which is known as the $f(\mathcal{R},L_m)$ theory of gravity. The corresponding Lagrangian can be modified by incorporating an analytic function of $T_{jk}T^{jk}$ in this generalization procedure for the $f(\mathcal{R},L_m)$ theory. $f(\mathcal{R},T^2)$ gravity or energy-momentum squared gravity is the result of selecting the corresponding Lagrangian. In 2014, Katirci and Kavuk \cite{kat} originally put forward this theory, which allows the existence of a term in the action functional that is proportional to $T_{jk}T^{jk}$.
The $f(\mathcal{R},G)$-gravity theory \cite{EMOS10} was one of these modified theories. It was developed by changing the previous Ricci scalar $\mathcal{R}$ by a function of $\mathcal{R}$ and $G$, the Gauss-Bonnet invariant. The $f(\mathcal{R},T)$-gravity theory, discovered by Harko et al. \cite{HLNO11}, was another modified theory. This is an extension of $f\left(\mathcal{R}\right)$-gravity (\cite{BBHL07}, \cite{CNO18}) in which the trace $T$ of the EMT is directly linked to any arbitrary function of $\mathcal{R}$. Several functional forms of $f(\mathcal{R})$ were previously provided in the following works: (\cite{cap}
-\cite{kde1}) which shows that this modified theory has a number of cosmological applications. \par

The literature mentioned above makes it very clear that more attention needs to be paid to $f(\mathcal{R})$ gravity, and there are still a lot of unanswered questions. Inspired by the foregoing investigations, this article is focused to investigate $P\left(GR_{4}\right)$ GRW spacetime satisfying $f(\mathcal{R})$ gravity. The $f(\mathcal{R})$ gravity model, $f\left(\mathcal{R}\right)=\mathcal{R}-\mu \mathcal{R}_{c}\tanh (\frac{\mathcal{R}}{\mathcal{R}_{c}})$ \cite{shi} satisfies local gravity constraint \cite{am}. In this case, $f_{\mathcal{R}\mathcal{R}}>0$ if $\mathcal{R}>0$ and $f_{\mathcal{R}}>0$ if $\mu <1$. The condition $f_{\mathcal{R}\mathcal{R}}>0$ is required for the consistency of local gravity tests \cite{dol}, for the presence of the matter-dominated epoch \cite{am1} and for the stability of cosmological perturbations \cite{ys}. In this paper, we choose the model $f\left(\mathcal{R}\right)=\mathcal{R}-\mu \mathcal{R}_{c}\tanh (\frac{\mathcal{R}}{\mathcal{R}_{c}})$ in which $\mu$ and $\mathcal{R}_{c}$ are positive constants, which is chosen to explain different ECs. The level of theoretical variety introduced by these higher curvature components is not found in more basic theories like General Relativity. This variability may open up new avenues for understanding and predicting late-time cosmic acceleration. Higher curvature components in gravity theories may result in changed dynamics at late-times, which may enable more realistic modelling of the cosmic acceleration at late-times and providing a better fit to observational data. The inclusion of higher curvature terms may be inspired by their importance in the effective action of quantum gravity theories, like string theory \cite{st}. Examining their consequences for late-time cosmology can provide insight into the interaction between quantum physics and gravity. Investigating late-time cosmology in these structures offers a special chance to test basic physics outside of general relativity. Higher curvature terms may affect the behaviour of cosmic structures or affect cosmological perturbations, among other particular cosmological effects. Studying these implications can provide important insights into the fundamentals of gravity theory. These ideas might provide an alternate explanation for cosmic acceleration to dark energy, which could help to solve some of the long-standing cosmological problems. Higher curvature terms can have an impact on gravitational wave generation and propagation. Examining late-time cosmology in these theories may have consequences for gravitational wave signal detection and interpretation. \par

This article is structured as:
In Section 2, an example of a $P\left(GR_{4}\right)$ spacetime is illustrated. $P\left(GR_{4}\right)$ GRW spacetimes and $P\left(GR_{4}\right)$ PFS are investigated in next two Sections. Finally, a $P\left(GR_{4}\right)$ GRW spacetime in $f(\mathcal{R})$ gravity theory is considered.

\section{Example of a $P\left(GR_{4}\right)$ spacetime}
\noindent
Choose a Lorentzian metric $g$ on $\mathbb{R}^{4}$ described by 
\begin{equation}\label{2.1}
ds^{2}=g_{ij}dy^{i}dy^{j}=\left(dy^{1}\right)^{2}+\left(y^{1}\right)^{2}\left(dy^{2}\right)^{2}
+\left(y^{2}\right)^{2}\left(dy^{3}\right)^{2}-\left(dy^{4}\right)^{2},
\end{equation}
where $i,j=1,2,3,4$. Using \eqref{2.1}, we observe that the non-vanishing components of the metric tensor are
\begin{equation}\label{2.2}
g_{11}=1,\quad g_{22}=\left(y^{1}\right)^{2},\quad g_{33}=\left(y^{2}\right)^{2},\quad g_{44}=-1
\end{equation}
and the associated contravariant components are
\begin{equation}\label{2.3}
g^{11}=1,\quad g^{22}=\dfrac{1}{\left(y^{1}\right)^{2}}\,,\quad g^{33}=\dfrac{1}{\left(y^{2}\right)^{2}}\,,\quad g^{44}=-1.
\end{equation}
Utilizing equations \eqref{2.2} and \eqref{2.3}, here we determine the components (non-vanishing) of the Christoffel symbols, the curvature tensor and the Ricci tensor and they are
\begin{equation}\label{2.4}
\Gamma_{22}^{1}=-y^{1},\quad\Gamma_{33}^{2}=-\dfrac{y^{2}}{\left(y^{1}\right)^{2}}\,,\quad\Gamma_{12}^{2}
=\dfrac{1}{y^{1}}\,,\quad\Gamma_{23}^{3}=\dfrac{1}{y^{2}}\,,
\end{equation}
\begin{equation}\label{2.5}
\mathcal{R}_{1332}=-\dfrac{y^{2}}{y^{1}}\,,\quad\mathcal{R}_{12}=-\dfrac{1}{y^{1}y^{2}}
\end{equation}
and the symmetric properties lead to the other components.\par
The covariant derivative of non-vanishing Ricci tensor is written as
\begin{equation}\label{2.6}
\mathcal{R}_{12,1}=\dfrac{1}{y^{2}\left(y^{1}\right)^{2}}\quad\mathrm{and}\quad\mathcal{R}_{12,2}
=\dfrac{1}{y^{1}\left(y^{2}\right)^{2}}\,.
\end{equation}
The one-forms we select are as follows:
\begin{equation}\label{2.7}
u_{i}\left( y\right)  =
\begin{cases}
1, & $ when $ i = 4\\
0, &  $ otherwise, $
\end{cases}
\end{equation}
\begin{equation}\label{2.8}
v_{i}\left( y\right)  =
\begin{cases}
y^{1}, & $ when $ i=2\\
y^{2}, &  $ when $ i=3 \\
\,0, &  $ otherwise $
\end{cases}
\end{equation}
and
\begin{equation}\label{2.9}
w_{i}\left( y\right)  =
\begin{cases}
y^{2}, & $ when $ i=1\\
y^{1}, &  $ when $ i=2 \\
\,0, &  $ otherwise $
\end{cases}
\end{equation}
for all $y\in\mathbb{R}^{4}$.\par

We consider $\mathcal{D}_{ij}$ as follows:
\begin{equation}\label{2.10}
\mathcal{D}_{ij}\left( y\right)  =
\begin{cases}
\dfrac{1}{\left(y^{1}\right)^{2}\left(y^{2}\right)^{2}}\,, & $ when $ i = 1,\, j=2\\
\quad\quad0, &  $ otherwise $
\end{cases}
\end{equation}
for all $y\in\mathbb{R}^{4}$. It is enough to examine at the subsequent equations in order to confirm the relation \eqref{1.7}:
\begin{equation}\label{2.11}
\mathcal{R}_{12,1}=u_{1}\mathcal{R}_{12}+v_{1}g_{12}+w_{1}\mathcal{D}_{12}
\end{equation}
and
\begin{equation}\label{2.12}
\mathcal{R}_{12,2}=u_{2}\mathcal{R}_{12}+v_{2}g_{12}+w_{2}\mathcal{D}_{12}.
\end{equation}
The other cases hold trivially.
\begin{align*}
\text{Now, right hand side of \eqref{2.11}} &=u_{1}\mathcal{R}_{12}+v_{1}g_{12}+w_{1}\mathcal{D}_{12}\\
&=0\cdot\left(\dfrac{-1}{y^{1}y^{2}}\right)+0+y^{2}\cdot\dfrac{1}{\left(y^{1}\right)^{2}\left(y^{2}\right)^{2}}\\
&=\dfrac{1}{y^{2}\left(y^{1}\right)^{2}}=\mathcal{R}_{12,1}.
\end{align*}
By applying the same deduction, it is possible to demonstrate that equation \eqref{2.12} is likewise true.\par
By virtue of \eqref{2.3} and \eqref{2.7} we find
\begin{equation}\label{2.13}
g^{ij}u_{i}u_{j}=-1,\quad\mathrm{and}\quad u^{i}\left(y\right)=g^{ij}u_{j}\left(y\right)=
\begin{cases}
	-1, & $ when $ i = 4\\
	\;\;0, &  $ otherwise $
\end{cases}
\end{equation}
Equations \eqref{2.10} and \eqref{2.13} together yield
\begin{equation}\label{2.14}
\mathcal{D}_{ij}u^{i}=0.
\end{equation}
Clearly, the trace$\left(\mathcal{D}_{ij}\right)=0.$ Therefore, $\left(\mathbb{R}^{4},g\right)$ is a $P\left(GR_{4}\right)$ spacetime.

\section{$P\left(GR_{4}\right)$ GRW spacetimes}
\noindent
{\bf Theorem A.} \cite{survey} A $\mathrm{M}^{4}$ is a GRW spacetime if and only if it admits a unit torse-forming time-like vector $u_{i}$:
\begin{equation}\label{4.1}
\nabla_{k}u_{i}=\varphi\left\{g_{ki}+u_{k}u_{i}\right\}
\end{equation}
and $u_{i}$ is an eigenvector of $\mathcal{R}_{ij}$, that is,
\begin{equation}\label{4.2}
\mathcal{R}_{ij}u^{j}=\theta u_{j}
\end{equation}
in which $\varphi$ and $\theta$ are non-zero scalars.\par
Multiplying \eqref{1.7} with $u^{i}$ and using \eqref{1.8} infers
\begin{equation}\label{4.3}
\left(\nabla_{k}\mathcal{R}_{ij}\right)u^{i}=u_{k}\mathcal{R}_{ij}u^{i}+v_{k}u_{j}.
\end{equation}
Since $\nabla_{k}\left(\mathcal{R}_{ij}u^{i}\right)=\left(\nabla_{k}\mathcal{R}_{ij}\right)u^{i}
+\mathcal{R}_{ij}\left(\nabla_{k}u^{i}\right)$, \eqref{4.3} becomes
\begin{equation}\label{4.4}
\nabla_{k}\left(\mathcal{R}_{ij}u^{i}\right)-\mathcal{R}_{ij}\left(\nabla_{k}u^{i}\right)
=u_{k}\mathcal{R}_{ij}u^{i}+v_{k}u_{j}.
\end{equation}
Equations \eqref{4.1}, \eqref{4.2} and \eqref{4.4} together imply
\begin{equation}\label{4.5}
\varphi\mathcal{R}_{jk}=\theta\varphi g_{jk}+\theta_{k}u_{j}-\theta u_{k}u_{j}-v_{k}u_{j}.
\end{equation}
Interchanging $j$ and $k$ in \eqref{4.5}, we obtain
\begin{equation}\label{4.6}
\varphi\mathcal{R}_{kj}=\theta\varphi g_{kj}+\theta_{j}u_{k}-\theta u_{j}u_{k}-v_{j}u_{k}.
\end{equation}
Subtracting \eqref{4.6} from \eqref{4.5}, we have
\begin{equation}\label{4.7}
\theta_{k}u_{j}=\theta_{j}u_{k}+v_{k}u_{j}-v_{j}u_{k}.
\end{equation}
Multiplying \eqref{4.7} with $u^{j}$, we acquire
\begin{equation}\label{4.8}
\theta_{k}=\left(v_{j}u^{j}\right)u_{k}-\left(\theta_{j}u^{j}\right)u_{k}+v_{k},
\end{equation}
that is,
\begin{equation}\label{4.9}
\theta_{k}=\left(f_{1}-f_{3}\right)u_{k}+v_{k},\quad\mathrm{where}\quad f_{1}=v_{j}u^{j}\quad\mathrm{and}\quad f_{3}=\theta_{j}u^{j}.
\end{equation}
From \eqref{4.5} and \eqref{4.9}, it follows that
\begin{equation}\label{4.10}
\mathcal{R}_{jk}=\theta g_{jk}+(\dfrac{f_{1}-f_{3}-\theta}{\varphi})u_{j}u_{k}.
\end{equation}
Hence, we reach:
\begin{theo}
A $P\left(GR_{4}\right)$ $\mathrm{GRW}$ spacetime represents a $\mathrm{PFS}$.
\end{theo}
\noindent
In light of equations \eqref{1.1}, \eqref{1.4} and \eqref{4.10}, we acquire
\begin{equation}\label{4.11}
\kappa\left(\dfrac{\sigma-p}{2}\right)=\theta
\end{equation}
and
\begin{equation}\label{4.12}
\kappa\left(\sigma+p\right)=\dfrac{f_{1}-f_{3}-\theta}{\varphi}\,.
\end{equation}
Equations \eqref{4.11} and \eqref{4.12} together give
\begin{equation}\label{4.13}
\dfrac{p}{\sigma}=\dfrac{f_{1}-f_{3}-\theta-2\theta\varphi}{f_{1}-f_{3}-\theta+2\theta\varphi}\,.
\end{equation}
We observe that \eqref{4.13} implies $p=0$ for $f_{1}=f_{3}+\theta\left(1+2\varphi\right)$, $\sigma=3p$ for $f_{1}=f_{3}+\theta\left(1+4\varphi\right)$ and $\sigma+p=0$ for $f_{1}=f_{3}+\theta$, respectively. Hence, we obtain:
\begin{cor}
A $ P\left(GR_{4}\right)$ $\mathrm{GRW}$ spacetime represents a
\begin{enumerate}
\item state equation of the form \eqref{4.13},
\item dust matter fluid for $f_{1}=f_{3}+\theta\left(1+2\varphi\right)$,
\item radiation era for $f_{1}=f_{3}+\theta\left(1+4\varphi\right)$ and
\item dark energy epoch of the Universe for $f_{1}=f_{3}+\theta$.
\end{enumerate}
\end{cor}	
\begin{defi}
A vector field $u_k$ is Riemann compatible \cite{Mantica3} if and only if it is Weyl compatible and
\begin{align}\label{4.14}
u_{i}\mathcal{R}^{h}_{j}u_{h}-u_{j}\mathcal{R}^{h}_{i}u_{h}=0.
\end{align}
\end{defi}

In a $P\left(GR_{4}\right)$ GRW spacetime

\begin{align}\label{4.15}
\mathcal{R}_{ij}={\alpha}g_{ij}+{\beta}u_{i}u_{j}
\end{align}	
in which
\begin{align}\label{4.16}
\alpha =\theta,
~~\beta=\frac{f_1-f_3-\theta}{\varphi}.
\end{align}	
 Multiplying equation \eqref{4.15} by $g^{il}$, we acquire

\begin{align}\label{4.17}
u_{i}\mathcal{R}^{h}_{j}u_{h}-u_{j}\mathcal{R}^{h}_{i}u_{h}= u_i(\alpha \delta^h_j+\beta u^hu_j)u_h\nonumber  \\
-u_j(\alpha \delta^h_i+\beta u^hu_i)u_h \nonumber\\
=0.
\end{align}

Thus the equation \eqref{4.14} is verified. Hence, we write:

\begin{theo}
  Every vector field of a $P\left(GR_{4}\right)$ GRW spacetime is Riemann compatible.
\end{theo}

The magnetic and electric parts of the Weyl tensor are given by
\begin{align}\label{4.18}
H_{ij}=u^ku^l\tilde{C}_{kijl},\;\;\;\;E_{ij}=u^ku^lC_{kijl},
\end{align}	
in which $u_ku^k=-1$ and $\tilde{C}_{kijl}=\frac{1}{2}{\epsilon}_{kilm}C^{lm}_{jl}$	
is the dual \cite{Bert}. These two tensors are traceless, symmetric and obey $E_{hk}u^h=0$ and $H_{hk}u^h=0$.\par
For $n=4$, from the definition of Weyl tensor, we provide
\begin{align}\label{4.19}
E_{ij}= \mathcal{R}_{hijk}u^hu^k-\frac{1}{2}(g_{ij}\mathcal{R}_{hk}-g_{ik}\mathcal{R}_{hj}+g_{hk}\mathcal{R}_{ij}-g_{hj}\mathcal{R}_{ik})u^hu^k\nonumber\\
+\frac{\mathcal{R}}{6}(g_{ij}g_{hk}-g_{ik}g_{hj}).
\end{align}
By hypothesis $P\left(GR_{4}\right)$ is a GRW spacetime.\par
Hence, using equation \eqref{4.15} in \eqref{4.19}, we infer
\begin{align}\label{4.20}
E_{ij}= \mathcal{R}_{hijk}u^hu^k+\frac{\alpha-\beta}{3}(g_{ij}+u_iu_j).
\end{align}
From equation \eqref{4.1}, we obtain
\begin{eqnarray}\label{4.21}
  \nabla_{k}\nabla_{j}u_{i} &=& \varphi_{k}\left\{g_{ij}+u_{i}u_{j}\right\}\nonumber\\&&
  \varphi [\varphi\left\{g_{ik}+u_{i}u_{k}\right\}u_j+\varphi\left\{g_{jk}+u_{j}u_{k}\right\}u_i]
\end{eqnarray}
and hence we get
\begin{equation}\label{4.22}
  \nabla_{k}\nabla_{j}u_{i}-\nabla_{j}\nabla_{k}u_{i}=\varphi_{k}\left\{g_{ij}+u_{i}u_{j}\right\}
  -\varphi_{j}\left\{g_{ik}+u_{i}u_{k}\right\}.
\end{equation}
From the above, we can easily acquire
\begin{equation}
  u_{h}\mathcal{R}^h_ijk=\varphi_{k}\left\{g_{ij}+u_{i}u_{j}\right\}
  -\varphi_{j}\left\{g_{ik}+u_{i}u_{k}\right\}\nonumber
\end{equation}
which entails
\begin{equation}\label{4.23}
  u^{h}\mathcal{R}_hijk=\varphi_{k}\left\{g_{ij}+u_{i}u_{j}\right\}
  -\varphi_{j}\left\{g_{ik}+u_{i}u_{k}\right\}
\end{equation}
Multiplying the foregoing equation by $u^k$ yields
\begin{equation}\label{4.24}
  u^{k}u^{h}\mathcal{R}_hijk=f\left\{g_{ij}+u_{i}u_{j}\right\},
\end{equation}
in which $f=u^k\varphi_{k}$.\par
Using equation \eqref{4.24} in \eqref{4.20}, we get
\begin{align}\label{4.25}
E_{ij}= \{f+\frac{\alpha-\beta}{3}\}(g_{ij}+u_iu_j).
\end{align}
Now, from equation \eqref{4.24}, we provide
\begin{equation}\label{4.26}
  u^{k}u^{h}\mathcal{R}_hk=3f.
\end{equation}
Also, from equation \eqref{4.15}, we acquire
\begin{equation}\label{4.27}
  u^{k}u^{h}\mathcal{R}_hk=-\frac{\alpha-\beta}{3}.
\end{equation}
Hence, the last two equations jointly give
\begin{equation}\label{4.28}
  f=-\frac{\alpha-\beta}{3}.
\end{equation}
Using equation \eqref{4.28} in \eqref{4.25}, we get $E_{ij}=0$.\par
Thus we can write
\begin{theo}
  In a $P\left(GR_{4}\right)$ GRW spacetime, the electric part of the spacetime vanishes.
\end{theo}

Let $P\left(GR_{4}\right)$ spacetime be a GRW spacetime. From the last two Theorems it follows that the velocity vector is Riemann compatible and the electric part of the spacetime under consideration vanishes. Hence, using Definition 1, we conclude that the velocity vector is Weyl compatible.\par

Also, we are aware that $H=0$, or the magnetic component of the Weyl tensor disappears when a spacetime's vector field is Weyl compatible \cite{Mantica3}. Therefore, the spacetime under consideration becomes conformally flat since $E=0$ and $H=0$ and hence is of petrov type O. Therefore we state:
\begin{theo}
  A $P\left(GR_{4}\right)$ GRW spacetime is conformally flat and is of petrov type O.
\end{theo}

\section{$P\left(GR_{4}\right)$ $\mathrm{perfect\, fluid\, spacetimes}$}
\noindent
Considering the covariant derivative of \eqref{1.1}, we infer that
\begin{equation}\label{5.1}
\nabla_{k}\mathcal{R}_{ij}=\alpha_{k}g_{ij}+\beta_{k}u_{i}u_{j}
+\beta\left\{u_{i}\left(\nabla_{k}u_{j}\right)+u_{j}\left(\nabla_{k}u_{i}\right)\right\}.
\end{equation}
Using \eqref{1.1} and \eqref{5.1} in \eqref{1.7}, we find
\begin{equation}\label{5.2}
\beta\left\{u_{i}\left(\nabla_{k}u_{j}\right)+u_{j}\left(\nabla_{k}u_{i}\right)\right\}=\left(\alpha u_{k}+v_{k}-\alpha_{k}\right)g_{ij}+\left(\beta u_{k}-\beta_{k}\right)u_{i}u_{j}+w_{k}\mathcal{D}_{ij}.
\end{equation}
Since $u_{j}u^{j}=-1$, therefore $u^{j}(\nabla_{k}u_{j})+u_{j}(\nabla_{k}u^{j})=0$, which implies
\begin{equation}\label{5.3}
u^{j}\left(\nabla_{k}u_{j}\right)=0.
\end{equation}
Multiplying \eqref{5.2} with $u^{i}$ and using \eqref{5.3}, we notice that
\begin{equation}\label{5.4}
\beta\left(\nabla_{k}u_{j}\right)=p_{k}u_{j},\quad\mathrm{where}\quad p_{k}=\beta u_{k}-\beta_{k}-\alpha u_{k}-v_{k}+\alpha_{k}.
\end{equation}
Again, multiplying \eqref{5.4} with $u^{j}$ and using \eqref{5.3}, we acquire
\begin{equation}\label{5.5}
p_{k}=0.
\end{equation}
Equations \eqref{5.4} and \eqref{5.5} reflect that
\begin{equation}\label{5.6}
\beta\left(\nabla_{k}u_{j}\right)=0,
\end{equation}
which entails the following cases: 
\vskip.1in
\noindent
{\bf Case 1.} For $\beta=0$, equation \eqref{1.4} gives us $\sigma+p=0$. Thus, the spacetime represents a dark energy epoch of the Universe.
\vskip.05in
\noindent
{\bf Case 2.} For $\beta$$\neq0$, \eqref{5.6} becomes
\begin{equation}\label{5.7}
\nabla_{k}u_{j}=0.
\end{equation}
Contracting with $g^{kj}$ yields
\begin{equation}\label{5.8}
\nabla_{k}u^{k}=0.
\end{equation}
Again multiplying equation \eqref{5.7} by $u^{k}$ gives
\begin{equation}\label{5.9}
\dot{u_{j}}=u^{k}\nabla_{k}u_{j}=0.
\end{equation}
The equations \eqref{5.7}-\eqref{5.9} show that the vector field $u_{k}$ is parallel, conservative,
and acceleration-free.\par
The vorticity tensor $\mu _{lk}$ is described by
\begin{equation}\label{c5}
  \mu _{lk}=\frac{1}{2}(\nabla_{l}a_{k}-\nabla_{k}a_{l})+\frac{1}{2}(a_{l}\dot{a_{k}}-a_{k}\dot{a_{l}}).
\end{equation}
Therefore, using equations \eqref{5.7} and \eqref{5.9}, we provide $\mu _{lk}=0$, that is, vorticity free.\par
The covariant gradient $\nabla_{k}u_{j}$ in a spacetime has the following standard decomposition \cite{HE73}:
\begin{equation}\label{5.10}
  \nabla_{k}u_{j}=\frac{\nabla_{i}u^{i}}{n-1}(g_{kj}+u_ku_j)-u_k\dot{u_{j}}+\mu_{kj}+\sigma_{kj}.
\end{equation}
In view of the above equations, the last equation provides $\sigma_{kj}=0$, that is, shear free.\par

A spacetime is said to be stationary if $u_{i}$ is Killing and static (\cite{S05}, \cite{SKMHH09}, p. 283) for irrotational vector $u_{i}$. For a smooth vector $v$,
\begin{equation*}
\pounds_{v}g_{lk}=\nabla_{l}v_{k}+\nabla_{k}v_{l},
\end{equation*}
$\pounds$ denotes the Lie derivative. As $\nabla_{k}u_{i}=0$, then $\pounds_{u}g_{ki}=0$, means that $u_{i}$ is Killing. Moreover, $\nabla_{k}u_{i}=0$ gives $u_{i}$ is irrotational. Therefore, the spacetime is static.\par
Hence, it is possible to declare the following:
\begin{theo}
A $P\left(GR_{4}\right)$ $\mathrm{PFS}$ represents either a dark energy epoch of the Universe or, the vector field $u_k$
is parallel, conservative, acceleration-free, vorticity-free, and shear-free and becomes a static spacetime.
\end{theo}

\section{$P\left(GR_{4}\right)$ GRW spacetime in $f(\mathcal{R})$ gravity}
Here, we will look into a few characteristics of a $P\left(GR_{4}\right)$ GRW spacetime in $f(\mathcal{R})$ gravity.\par
Choose the Einstein-Hilbert action
\begin{align}
S=\frac{1}{2k^2}\int{d^4x{\sqrt{-g}}f(\mathcal{R})}+\int{d^4x{\sqrt{-g}}L_{m}}
\nonumber
\end{align}
in which $L_m$ is the matter Lagrangian density described as
\begin{align}
T_{ij}=-\frac{2}{\sqrt{-g}}\frac{\delta (\sqrt{-g}L_m)}{\delta g^{ij}}
\nonumber
\end{align}
In this case, $\kappa^{2}=8{\pi}G$, $G$ stands for the Newton's constant and hence the modified formula can be written as \cite{Sotiriou}
\begin{align}
S=\frac{1}{2k^2}\int{d^4x{\sqrt{-g}}f(\mathcal{R})}.
\nonumber
\end{align}

The field equations can be expressed in the following form by applying the variation with $g^{ij}$
\begin{align}\label{36}
f_{\mathcal{R}}(\mathcal{R})\mathcal{R}_{ij}-\frac{f(\mathcal{R})}{2}g_{ij}+(g_{ij}{\square}
-\nabla_{i}\nabla_{j})f_{\mathcal{R}}(\mathcal{R})=k^2T_{ij}
\end{align}
where 
$\square$ indicates the D'Alembert's operator.\par
Taking trace of equation \eqref{36} provides
\begin{align}\label{37}
3{\square}f_{\mathcal{R}}(\mathcal{R})+\mathcal{R}f_{\mathcal{R}}(\mathcal{R})-2f(\mathcal{R})=k^2T.
\end{align}	

Subtracting the expression $\frac{\mathcal{R}f_{\mathcal{R}}(\mathcal{R})}{2} g_{ij}$ from \eqref{36}, we infer
\begin{align}\label{38}
f_{\mathcal{R}}(\mathcal{R})\mathcal{R}_{ij}-\frac{\mathcal{R}f_{\mathcal{R}}(\mathcal{R})}{2}g_{ij}
=k^2T_{ij}+k^2T^{(curve)}_{ij}
\end{align}
such that
\begin{align}\label{39}
T^{(eff)}_{ij}=T_{ij}+T^{(curve)}_{ij}
\end{align}
in which
\begin{align}\label{40}
T^{(curve)}_{ij}=\frac{1}{k^2}\big[\frac{(f(\mathcal{R})-\mathcal{R}f_{\mathcal{R}}(\mathcal{R}))}{2}g_{ij}
+(\nabla_{i}\nabla_{j}-g_{ij}{\square})f_{\mathcal{R}}(\mathcal{R})\big].
\end{align}

Using the equation \eqref{38}, we acquire

\begin{align}\label{41}
\mathcal{R}_{ij}-\frac{\mathcal{R}}{2}g_{ij}=\frac{k^2}{f_{\mathcal{R}}(\mathcal{R})}T_{ij}^{(eff)}
\end{align}	
which obeys the equations \eqref{39} and  \eqref{40}.\par

Multiplying equation \eqref{4.10} with $g^{jk}$, we reach
\begin{equation}\label{ee5}
\dfrac{f_{1}-f_{3}-\theta}{\varphi}=4\theta-\mathcal{R}.
\end{equation}
Utilizing \eqref{ee5} in \eqref{4.10}, we get
\begin{equation}\label{eee5}
\mathcal{R}_{jk}=\theta g_{jk}+\left(4\theta-\mathcal{R}\right)u_{j}u_{k}.
\end{equation}
Taking covariant derivative of the foregoing equation yields
\begin{eqnarray}\label{1z}
  \nabla_{i}\mathcal{R}_{jk} &=& \theta_{i} g_{jk}+\left(4\theta_{i}-\nabla_{i}\mathcal{R}\right)u_{j}u_{k}\nonumber\\&&
  +\left(4\theta-\mathcal{R}\right)[u_{k}\nabla_{i}u_{j}+u_{j}\nabla_{i}u_{k}].
\end{eqnarray}
Multiplying the above equation by $u^{k}$ provides
\begin{equation}\label{2z}
  \left(4\theta-\mathcal{R}\right)\nabla_{i}u_{j}=(\nabla_{i}\mathcal{R}-3\theta_{i})u_{j}
  -u^{k}\nabla_{i}\mathcal{R}_{jk}.
\end{equation}
Again multiplying the previous equation by $u^{i}$ gives
\begin{equation}\label{3z}
  \left(4\theta-\mathcal{R}\right)\dot{u_{j}}=(\dot{\mathcal{R}}-3f_{3}+\theta-f_{1})u_{j},
\end{equation}
where $u^{i}\nabla_{i}u_{j}=\dot{u_{j}}$ and $u^{i}\nabla_{i}\mathcal{R}=\dot{\mathcal{R}}$.\par
Using $u_{j}\dot{u_{j}}=0$, we readily conclude that the acceleration vector
\begin{equation}\label{4z}
  \dot{u_{j}}=0.
\end{equation}
Now transvecting equation \eqref{1.9} by $u^{i}$, we acquire
\begin{equation}\label{5z}
  \dot{\mathcal{R}}+\mathcal{R}=4f_{1}.
\end{equation}
Covariant derivative of equation \eqref{1.9} produces
\begin{equation}\label{6z}
  \nabla_{j}\nabla_{i}\mathcal{R}=u_{i}(\nabla_{j}\mathcal{R})+\mathcal{R}(\nabla_{j}u_{i})+4\nabla_{j}v_{i}.
\end{equation}
Interchanging $i$ and $j$ yields
\begin{equation}\label{7z}
\nabla_{i}\nabla_{j}\mathcal{R}=u_{j}(\nabla_{i}\mathcal{R})+\mathcal{R}(\nabla_{i}u_{j})+4\nabla_{i}v_{j}.
\end{equation}
Utilizing the last two equations, we get
\begin{equation}\label{8z}
  \mathcal{R}[\nabla_{i}u_{j}-\nabla_{j}u_{i}]+[4v_{i}u_{j}-4u_{i}v_{j}]+4[\nabla_{i}v_{j}-\nabla_{j}v_{i}]
\end{equation}
Transvecting the last equation by $u^{i}$, we obtain
\begin{equation}\label{9z}
  v_{j}=-f_{1}u_{j},
\end{equation}
since $\dot{u_{i}}=0$.\par
Suppose $v_{i}$ is parallel. Then using equation \eqref{9z} in equation \eqref{8z}, we provide
\begin{equation}\label{10z}
  \nabla_{i}\nabla_{j}\mathcal{R}=(\mathcal{R}-4f_{1})u_{i}u_{j}.
\end{equation}
Since $f(\mathcal{R})$ is an analytic function, then we write
\begin{equation}\label{11z}
\nabla_{i}\nabla_{j}f_{\mathcal{R}}(\mathcal{R})=f_{\mathcal{R}\mathcal{R}}(\mathcal{R})\nabla_{i}\nabla_{j}\mathcal{R}
+f_{\mathcal{R}\mathcal{R}\mathcal{R}}(\mathcal{R})(\nabla_{i}\mathcal{R})(\nabla_{j}\mathcal{R}).
\end{equation}
Hence, equation \eqref{11z} yields
\begin{equation}\label{12z}
  {\square}f_{\mathcal{R}}(\mathcal{R})=f_{\mathcal{R}\mathcal{R}}(\mathcal{R}){\square}\mathcal{R}
  +f_{\mathcal{R}\mathcal{R}\mathcal{R}}(\mathcal{R})g^{ij}(\nabla_{i}\mathcal{R})(\nabla_{j}\mathcal{R}).
\end{equation}
Now, using the equations \eqref{1.9}, \eqref{10z} and \eqref{11z}, we acquire
\begin{eqnarray}\label{13z}
  \nabla_{i}\nabla_{j}f_{\mathcal{R}}(\mathcal{R}) &=& \{(\mathcal{R}-4f_{1})f_{\mathcal{R}\mathcal{R}}(\mathcal{R})+(\mathcal{R}^{2}
  -8\mathcal{R}f_{1})f_{\mathcal{R}\mathcal{R}\mathcal{R}}(\mathcal{R})\}u_{i}u_{j}\nonumber\\&&
  +16f_{\mathcal{R}\mathcal{R}\mathcal{R}}(\mathcal{R})v_{i}v_{j}.
\end{eqnarray}
Multiplying equation \eqref{13z} by $g^{ij}$ and making use of the equations \eqref{10z} and \eqref{12z}, we get
\begin{equation}\label{14z}
{\square}f_{\mathcal{R}}(\mathcal{R})=(4f_{1}-\mathcal{R})f_{\mathcal{R}\mathcal{R}}(\mathcal{R})
-(8\mathcal{R}f_{1}-\mathcal{R}^{2}+16)f_{\mathcal{R}\mathcal{R}\mathcal{R}}(\mathcal{R}).
\end{equation}
Again, utilizing the equations \eqref{39}, \eqref{40} and \eqref{41}, we provide
\begin{align}\label{15z}
f_{\mathcal{R}}(\mathcal{R})(\mathcal{R}_{ij}-\frac{\mathcal{R}}{2}g_{ij})=k^2T_{ij}
+\frac{f(\mathcal{R})-\mathcal{R}f_{\mathcal{R}}(\mathcal{R})}{2}g_{ij}+(\nabla_{i}\nabla_{j}
-g_{ij}{\square})f_{\mathcal{R}}(\mathcal{R}).
\end{align}	
Substituting the equations \eqref{13z} and \eqref{14z} in equation \eqref{15z} infer
\begin{eqnarray}\label{16z}
&&f_{\mathcal{R}}(\mathcal{R})\big(\mathcal{R}_{ij}-\frac{\mathcal{R}}{2}g_{ij}\big)=\kappa^2T_{ij}\nonumber\\&&
+\big[\frac{f(\mathcal{R})-\mathcal{R}f_{\mathcal{R}}(\mathcal{R})}{2}-(8\mathcal{R}f_{1}-\mathcal{R}^{2}+16)f_{\mathcal{R}\mathcal{R}\mathcal{R}}(\mathcal{R})\nonumber\\&& -(4f_{1}-\mathcal{R})f_{\mathcal{R}\mathcal{R}}(\mathcal{R})\big]g_{ij}\nonumber\\&&
-\big[-(8\mathcal{R}f_{1}-\mathcal{R}^{2})f_{\mathcal{R}\mathcal{R}\mathcal{R}}(\mathcal{R}) +(4f_{1}-\mathcal{R})f_{\mathcal{R}\mathcal{R}}(\mathcal{R})\big]u_{i}u_{j}.
\end{eqnarray}	
Using equation \eqref{eee5} in equation \eqref{16z}, we acquire
\begin{eqnarray}\label{17z}
T_{ij}&=&-\frac{1}{\kappa^2}
\big[\frac{f(\mathcal{R})-2\theta f_{\mathcal{R}}(\mathcal{R})}{2}
-(8\mathcal{R}f_{1}-\mathcal{R}^{2}+16)f_{\mathcal{R}\mathcal{R}\mathcal{R}}(\mathcal{R})\nonumber\\&& -(4f_{1}-\mathcal{R})f_{\mathcal{R}\mathcal{R}}(\mathcal{R})\big]g_{ij}\nonumber\\&&
+\frac{1}{\kappa^2}\big[-(8\mathcal{R}f_{1}-\mathcal{R}^{2})f_{\mathcal{R}\mathcal{R}\mathcal{R}}(\mathcal{R})\nonumber\\&&
 +(4f_{1}-\mathcal{R})f_{\mathcal{R}\mathcal{R}}(\mathcal{R})-(4\theta-\mathcal{R})f_{\mathcal{R}}\mathcal{R}\big]u_{i}u_{j}.
\end{eqnarray}	
Therefore, using equations \eqref{1.2} and \eqref{17z}, we find
\begin{eqnarray}\label{18z}
p&=&-\frac{1}{\kappa^2}
\big[\frac{f(\mathcal{R})-2\theta f_{\mathcal{R}}(\mathcal{R})}{2}
-(8\mathcal{R}f_{1}-\mathcal{R}^{2}+16)f_{\mathcal{R}\mathcal{R}\mathcal{R}}(\mathcal{R})\nonumber\\&& -(4f_{1}-\mathcal{R})f_{\mathcal{R}\mathcal{R}}(\mathcal{R})\big]
\end{eqnarray}
and
\begin{eqnarray}\label{19z}
\sigma&=&\frac{1}{\kappa^2}
\big[\frac{f(\mathcal{R})-(10\theta-2\mathcal{R}) f_{\mathcal{R}}(\mathcal{R})}{2}\nonumber\\&&
-(16\mathcal{R}f_{1}-2\mathcal{R}^{2}+16)f_{\mathcal{R}\mathcal{R}\mathcal{R}}(\mathcal{R})\big].
\end{eqnarray}
Therefore, we provide:
\begin{theo}
In a $P\left(GR_{4}\right)$ GRW spacetime satisfying $f(\mathcal{R})$ gravity, $p$ and $\sigma$ are described by \eqref{18z} and \eqref{19z}, respectively.
\end{theo}
If $f(\mathcal{R})=\mathcal{R}$, then $f(\mathcal{R})$ theory reduces to Einstein theory. In this case the equations \eqref{18z} and \eqref{19z} takes the form
\begin{eqnarray}\label{20z}
p&=&-\frac{1}{\kappa^2}
[\frac{\mathcal{R}-2\theta }{2}]
\end{eqnarray}
and
\begin{eqnarray}\label{21z}
\sigma&=&\frac{1}{\kappa^2}
[\frac{3\mathcal{R}-10\theta}{2}].
\end{eqnarray}
Thus, we state:
\begin{cor}
In a $P\left(GR_{4}\right)$ GRW spacetime satisfying $f(\mathcal{R})$ gravity with the condition $f(\mathcal{R})=\mathcal{R}$, $p$ and $\sigma$ are given by \eqref{20z} and \eqref{21z}, respectively.
\end{cor}
Equations \eqref{20z} and \eqref{21z} jointly yield
\begin{equation}\label{22z}
\dfrac{p}{\sigma}=-\dfrac{\mathcal{R}-2\theta}{3\mathcal{R}-10\theta}.
\end{equation}
We observe that \eqref{22z} implies $p=0$ for $\mathcal{R}=2\theta$, $\sigma=3p$ for $\mathcal{R}=\frac{8}{3}\theta$ and $\sigma+p=0$ for $\mathcal{R}=4\theta$, respectively. Hence, we write:
\begin{rem}
A $P\left(GR_{4}\right)$ GRW spacetime satisfying $f(\mathcal{R})$ gravity with the condition $f(\mathcal{R})=\mathcal{R}$, represents a
\begin{enumerate}
\item state equation of the form \eqref{22z},
\item dust matter fluid for $\mathcal{R}=2\theta$,
\item radiation era for $\mathcal{R}=\frac{8}{3}\theta$ and
\item dark energy epoch of the Universe for $\mathcal{R}=4\theta$.
\end{enumerate}
\end{rem}

\subsection{Energy Conditions}
In the following subsection, we verify the ECs for the $f\left(\mathcal{R}\right)$-gravity model $f\left(\mathcal{R}\right)=\mathcal{R}-\mu \mathcal{R}_{c}\tanh (\frac{\mathcal{R}}{\mathcal{R}_{c}})$ \cite{shi}, in which $\mu$ and $\mathcal{R}_{c}$ are positive constants. \par

\noindent
In modified gravity the ECs are demonstrated as
\begin{align*}
 \mathrm{NEC}&\quad\mathrm{if\;and\;only\;if}\quad\sigma+p\geq0,\\
 \mathrm{TEC}&\quad\mathrm{if\;and\;only\;if}\quad\sigma-3p\geq0,\\
 \mathrm{SEC}&\quad\mathrm{if\;and\;only\;if}\quad\sigma+p\geq0\quad\mathrm{and}\quad\sigma+3p\geq0,\\
 \mathrm{DEC}&\quad\mathrm{if\;and\;only\;if}\quad\sigma\pm p\geq0\quad\mathrm{and}\quad\sigma\geq0,\\
 \mathrm{WEC}&\quad\mathrm{if\;and\;only\;if}\quad\sigma+p\geq0\quad\mathrm{and}\quad\sigma\geq0.
\end{align*}
With the help of \eqref{18z} and \eqref{19z}, $\sigma$ and $p$ are given by
\begin{align}\label{5.11}
\kappa\sigma=& \frac{\mathcal{R}}{2}-\frac{1}{2}\mu \mathcal{R}_{c}\tanh (\frac{\mathcal{R}}{\mathcal{R}_{c}})-(5\theta-\mathcal{R})[1-\mu(1-\tanh(\frac{\mathcal{R}}{\mathcal{R}_{c}})^{2})]
\nonumber\\& -(16\mathcal{R}f_{1}-2\mathcal{R}^{2}+16)
\big[\frac{2\mu(1-\tanh(\frac{\mathcal{R}}{\mathcal{R}_{c}})^{2})^{2}}{\mathcal{R}_{c}^{2}}
-\frac{4\mu \tanh(\frac{\mathcal{R}}{\mathcal{R}_{c}})^{2}
(1-\tanh(\frac{\mathcal{R}}{\mathcal{R}_{c}})^{2})}{\mathcal{R}_{c}^{2}}\big],
\end{align}
\begin{align}\label{5.12}
\kappa p=&- \frac{\mathcal{R}}{2}+\frac{1}{2}\mu \mathcal{R}_{c}\tanh (\frac{\mathcal{R}}{\mathcal{R}_{c}})+\theta[1-\mu(1-\tanh(\frac{\mathcal{R}}{\mathcal{R}_{c}})^{2})]
\nonumber\\& -(8\mathcal{R}f_{1}-\mathcal{R}^{2}+16)
\big[\frac{2\mu(1-\tanh(\frac{\mathcal{R}}{\mathcal{R}_{c}})^{2})^{2}}{\mathcal{R}_{c}^{2}}
-\frac{4\mu \tanh(\frac{\mathcal{R}}{\mathcal{R}_{c}})^{2}
(1-\tanh(\frac{\mathcal{R}}{\mathcal{R}_{c}})^{2})}{\mathcal{R}_{c}^{2}}\big]\nonumber\\&
-\frac{2(4f_{1}-\mathcal{R})\mu \tanh(\frac{\mathcal{R}}{\mathcal{R}_{c}})^{2}
(1-\tanh(\frac{\mathcal{R}}{\mathcal{R}_{c}})^{2})}{\mathcal{R}_{c}}.
\end{align}
The ECs for the above model are now examined. The ECs for this arrangement may now be discussed using equations \eqref{5.11} and \eqref{5.12}.\par
\begin{tabulary}{\linewidth}{CC}
	\includegraphics[height=0.24\textheight]{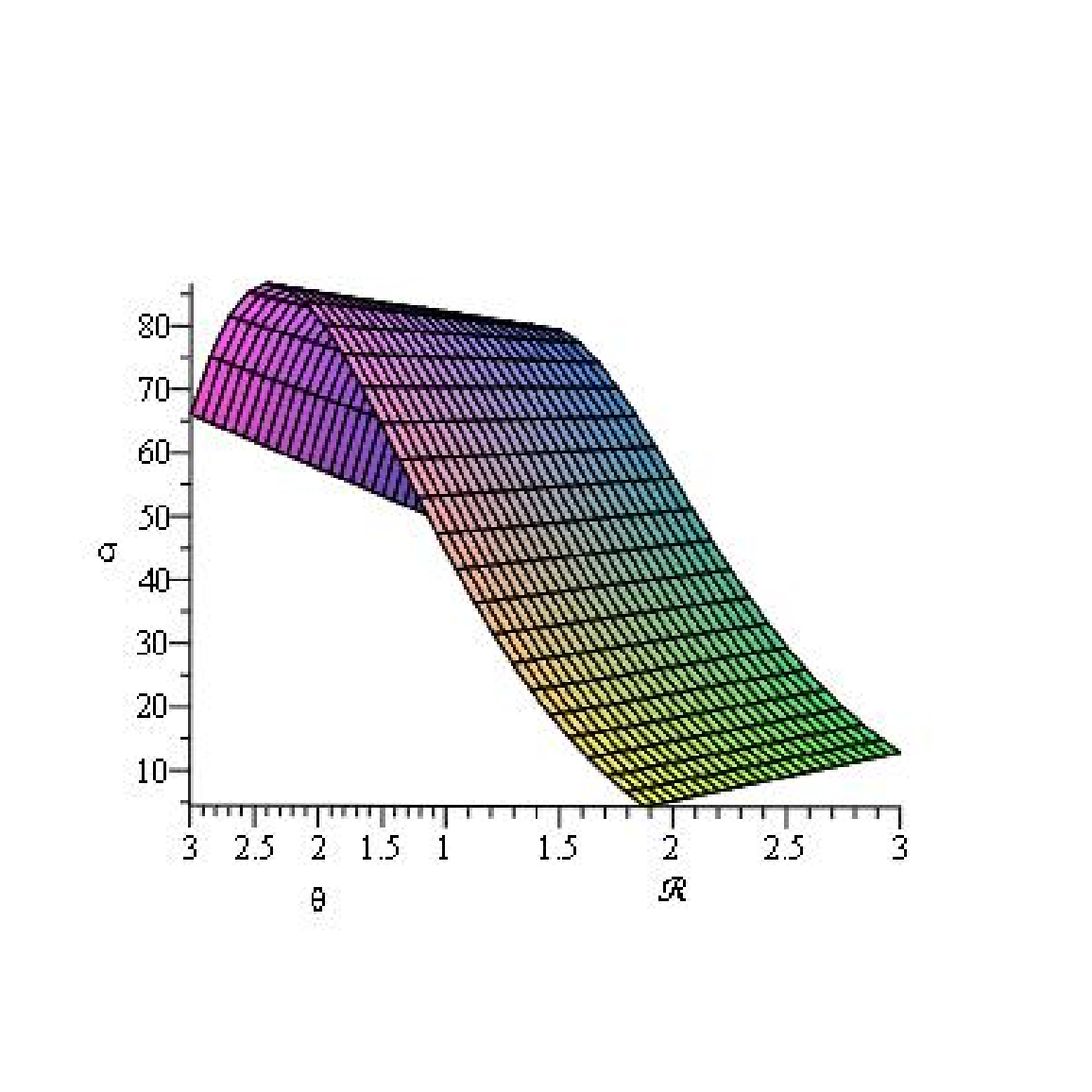}
	&
	\includegraphics[height=0.24\textheight]{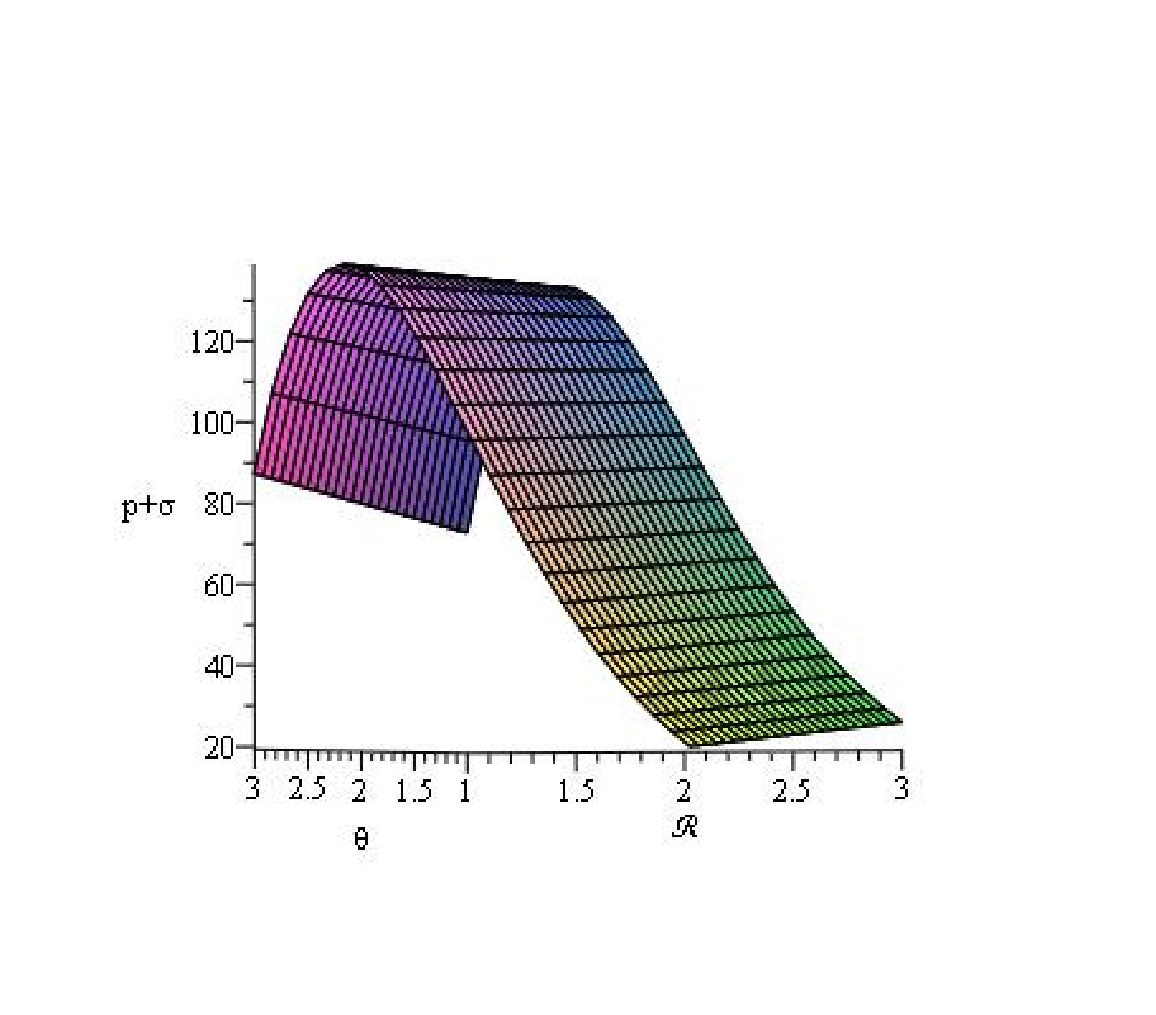}
	\\
	{\bf Fig. 1:} Development of $\sigma$ with reference to $\mathcal{R}$ and $\theta$ &{\bf Fig. 2:} Development of $p+\sigma$ with reference to $\mathcal{R}$ and $\theta$
	
\end{tabulary}
\begin{tabulary}{\linewidth}{CC}
	
	\includegraphics[height=0.25\textheight]{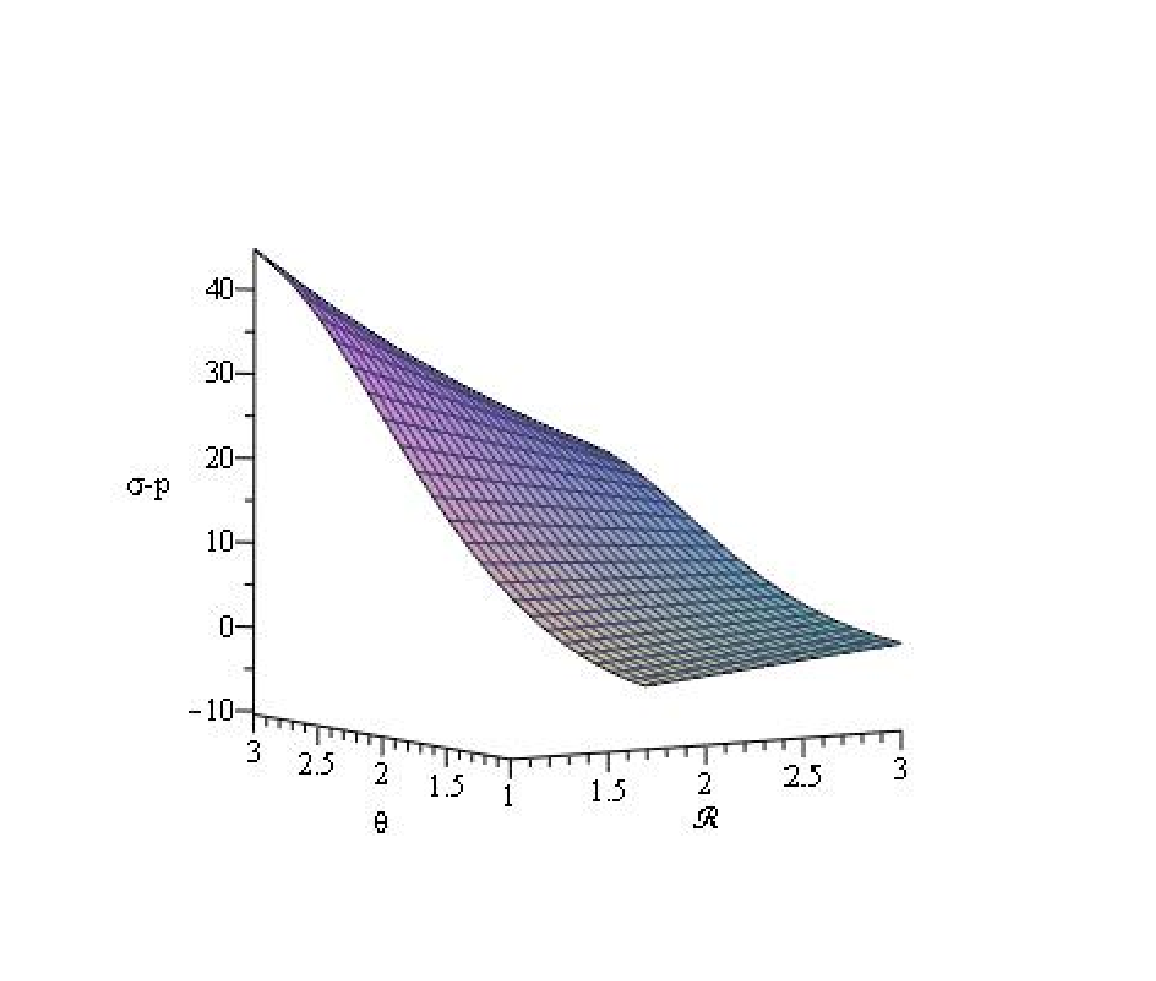}
    &
	\includegraphics[height=0.24\textheight]{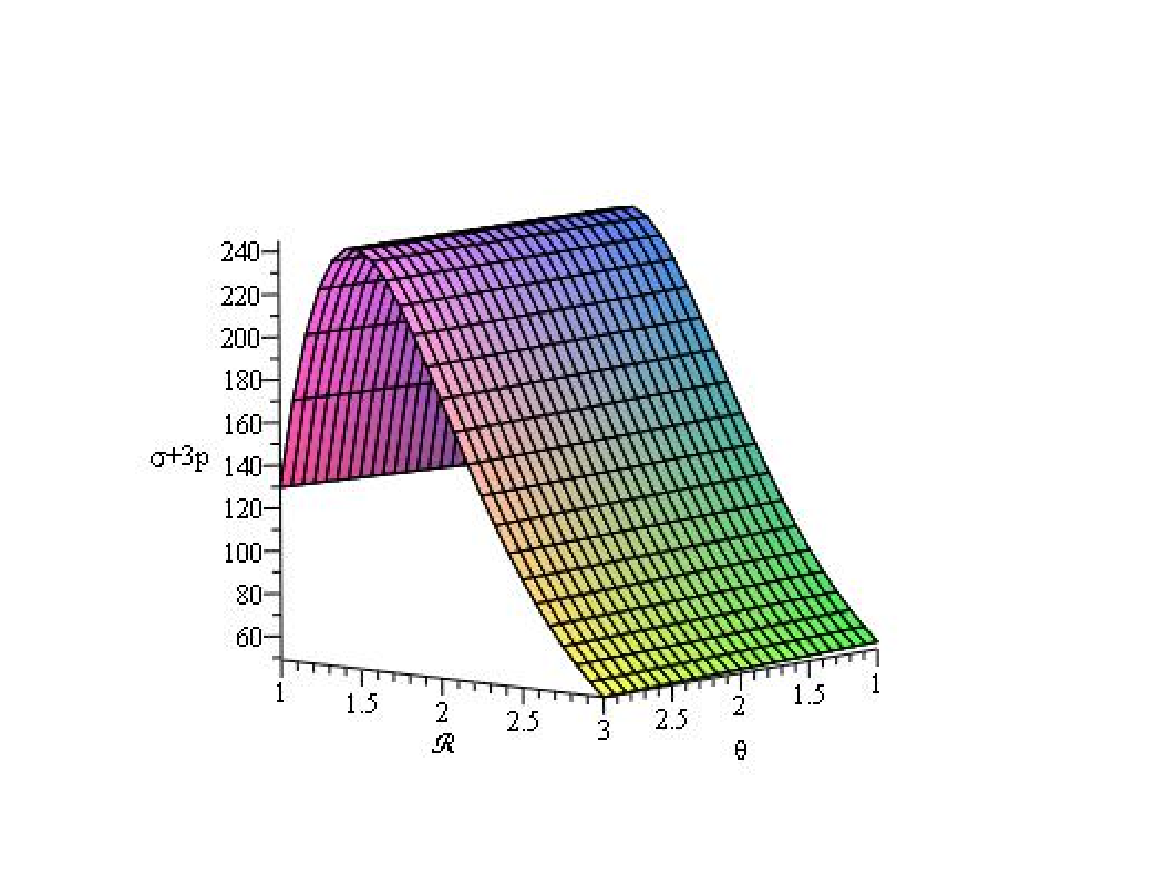}
    \\	
	
	{\bf Fig. 3:} Development of $\sigma-p$ with reference to $\mathcal{R}$ and $\theta$  &{\bf Fig. 4:} Development of $\sigma+3p$ with reference to $\mathcal{R}$ and $\theta$
	
\end{tabulary}

\begin{tabulary}{\linewidth}{CC}
	
	\includegraphics[height=0.25\textheight]{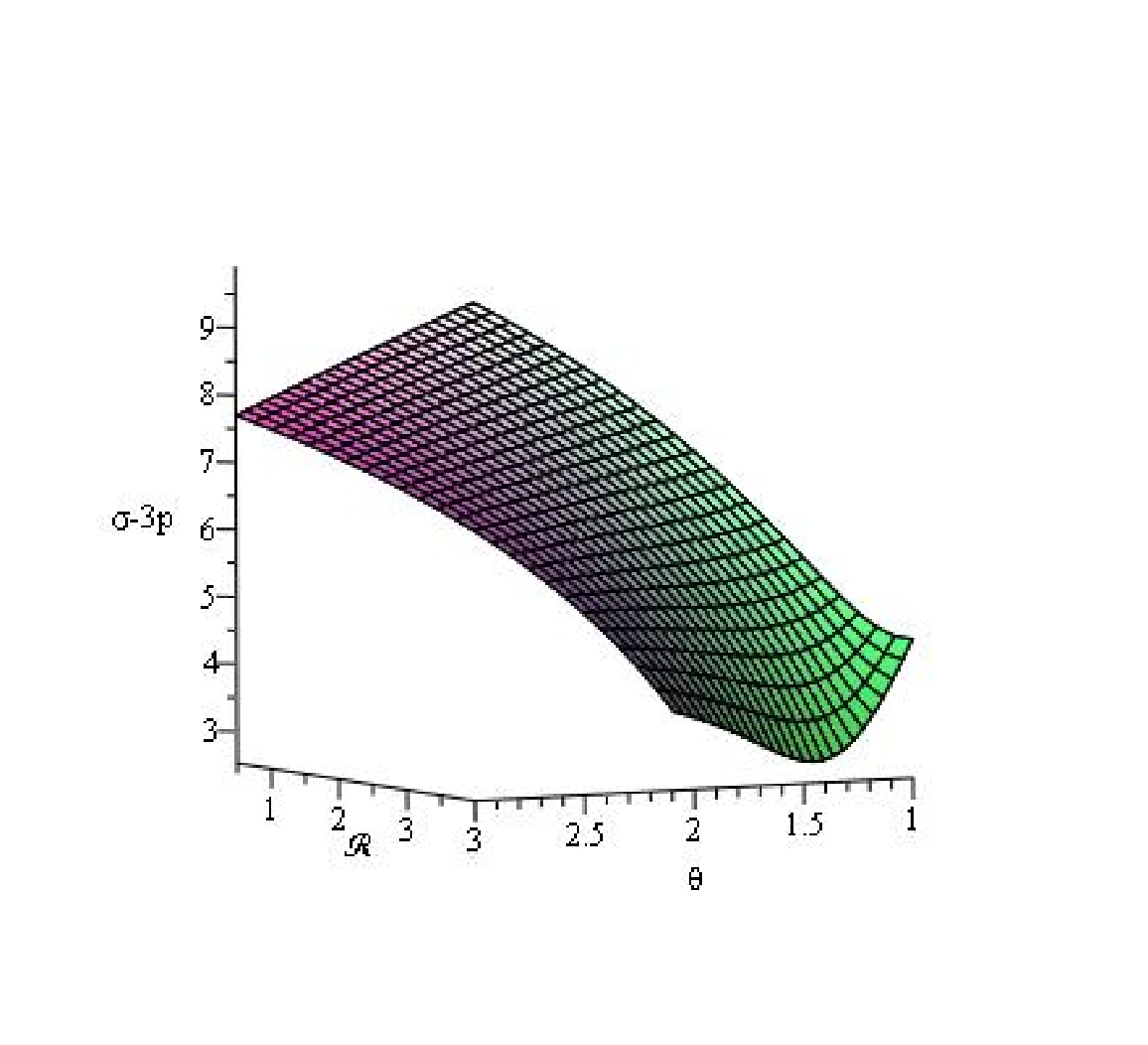}

	{\bf Fig. 5:} Development of $\sigma-3p$ with reference to $\mathcal{R}$ and $\theta$
	
\end{tabulary}

Figures $1$ and $2$ demonstrate that, for parameters $\theta, \mathcal{R}\in\left[1,3\right]$, the energy density and $p+\sigma$ can not be negative, and that, for larger values of $\theta$ and $\mathcal{R}$, they are high. Because NEC is a component of WEC, NEC and WEC are fulfilled. The $\sigma-\rho$ profile for $\theta, \mathcal{R}\in\left[1,3\right]$ is positive, as seen in Fig. $3$. It is evident from Figs. $1$, $2$, and $3$ that DEC is validated. Furthermore, we can observe that SEC is satisfied from Figs. $2$ and $4$, and this finding yields the late-time acceleration of the Cosmos\cite{LDMS21}. Moreover, each result aligns with the $\Lambda$CDM model \cite{AAAABBBBBBB20}. Fig. $5$ shows that TEC is satisfied.

\subsection{}
 The field equations of $f(\mathcal{R})$ gravity are as follows:
\begin{eqnarray}
  \kappa T_{lk} &=& f^{\prime}(\mathcal{R})\mathcal{R}_{lk}-f^{\prime\prime\prime}(\mathcal{R})\nabla_{l}\mathcal{R}\nabla_{k}\mathcal{R}
  -f^{\prime\prime}(\mathcal{R})\nabla_{l}\nabla_{k}\mathcal{R}\nonumber\\&&
  +g_{lk}[f^{\prime\prime\prime}(\mathcal{R})\nabla_{m}\mathcal{R}\nabla^{m}\mathcal{R}+f^{\prime\prime}(\mathcal{R})\nabla^{2}\mathcal{R}-\frac{1}{2}f(\mathcal{R})].
\end{eqnarray}
For $\mathcal{R}=$ constant, we acquire
\begin{equation}\label{d2}
 \mathcal{R}_{lk}-\frac{f}{2f^{\prime}}g_{lk}=\frac{\kappa}{f^{\prime}}T_{lk}.
\end{equation}
Using equation \eqref{1.2} in equation \eqref{d2}, we infer
\begin{equation}\label{d3}
 \mathcal{R}_{lk}=\frac{f}{2f^{\prime}}g_{lk}+\frac{\kappa}{f^{\prime}}[(\sigma+p)u_{k}u_{l}+p g_{kl}].
\end{equation}
Making use of equations \eqref{4.10} and \eqref{d3}, we obtain
\begin{equation}\label{d4}
 \frac{\kappa}{f^{\prime}}(\sigma+p)=\frac{f_1-f_3-\theta}{\varphi}
\end{equation}
and
\begin{equation}\label{d5}
\frac{p\kappa}{f^{\prime}}+ \frac{f}{2f^{\prime}}= \theta.
\end{equation}
Solving the last two equations, we get
\begin{equation}\label{d6}
  p=\frac{f^{\prime}\theta}{\kappa}-\frac{f}{2\kappa}
\end{equation}
and
\begin{equation}\label{d7}
  \sigma=-\frac{f^{\prime}(f_1-f_3-\theta-\theta\varphi)}{\kappa\varphi}+\frac{f}{2\kappa}.
\end{equation}
Therefore, we provide:
\begin{theo}
In a $P\left(GR_{4}\right)$ GRW spacetime satisfying $f(\mathcal{R})$ gravity with constant Ricci scalar, $p$ and $\sigma$ are described by \eqref{d6} and \eqref{d7}, respectively.
\end{theo}

\subsection{Energy Conditions}
In general relativity, energy conditions are vital tools to study black holes and wormholes in numerous modified gravities. To specify certain energy conditions in our current investigation of the $f(\mathcal{R})$ gravity, we must find the effective isotropic pressure $p^{eff}$ and the effective energy density $\sigma^{eff}$ (see, \cite{ade}-\cite{ade1}).\par

Equation \eqref{d2}, can be rewritten in the subsequent form
\begin{equation}\label{d8}
 \mathcal{R}_{lk}-\frac{\mathcal{R}}{2}g_{lk}=\frac{\kappa}{f^{\prime}}T_{lk}^{eff},
\end{equation}
in which
\begin{equation}\label{d9}
  T_{lk}^{eff}=T_{lk}+\frac{(f-\mathcal{R}f^{\prime})}{2\kappa}g_{lk}.
\end{equation}
Then equation \eqref{1.2} reduces to
\begin{equation}
\label{d10}
T_{kl}^{eff}=(\sigma^{eff}+p^{eff})u_{k}u_{l}+p^{eff} g_{kl},
\end{equation}
in which $p^{eff}=p+\frac{(f-\mathcal{R}f^{\prime})}{2\kappa}$ and $\sigma^{eff}=\sigma-\frac{(f-\mathcal{R}f^{\prime})}{2\kappa}$.\par
In our case, using equations \eqref{d6} and \eqref{d7}, we provide
\begin{equation}\label{d11}
  p^{eff}=\frac{f^{\prime}\theta}{\kappa}-\frac{(\mathcal{R}f^{\prime})}{2\kappa}
\end{equation}
and
\begin{equation}\label{d12}
  \sigma^{eff}=\frac{f^{\prime}(f_1-f_3-\theta-\theta\varphi)}{\kappa\varphi}+\frac{(\mathcal{R}f^{\prime})}{2\kappa}.
\end{equation}
Now we investigate the energy conditions in a $P\left(GR_{4}\right)$ GRW spacetime with non-zero constant Ricci scalar obeying $f(\mathcal{R})$ gravity. With the help of equations \eqref{d11} and \eqref{d12}, we find the NEC, TEC, DEC, WEC and SEC in this set up and they are given as follows:

\begin{table}[h]
    \centering
        \begin{tabular}{|c|c|c|}
    \hline
    \multicolumn{3}{|c|}{Table 1}\\
    \hline
     \multicolumn{3}{|c|}{Validity of Energy Conditions in $P\left(GR_{4}\right)$ GRW spacetime }\\
    \hline
      Energy Condition & Inequalities & Conditions of validation\\
        \hline
        NEC & $p^{eff}+\sigma^{eff} \geq 0$ & $ \mathcal{R} \leq (f_1-f_3-\theta)$\\
         & & \\
         \hline
        TEC &   $\sigma^{eff}-3p^{eff} \geq 0$ & $\mathcal{R}\leq \frac{(f_1-f_3-\theta-4\theta\varphi)}{\varphi}$\\
         & & \\
        \hline
        WEC & $\sigma^{eff} \geq 0 $ and  $p^{eff}+\sigma^{eff} \geq 0$ & $\mathcal{R}\leq \frac{(f_1-f_3-\theta-\theta\varphi)}{\varphi}$\\
         & & and $ \mathcal{R} \leq (f_1-f_3-\theta)$\\
        \hline
        DEC & $\sigma^{eff} \geq 0 $ and $\sigma^{eff} \pm p^{eff} \geq 0$ & $\frac{f_1-f_3}{\theta(1+2\varphi)}\geq 0$ , $\mathcal{R}\leq \frac{(f_1-f_3-\theta-\theta\varphi)}{\varphi}$\\
          & & and $ \mathcal{R} \leq (f_1-f_3-\theta)$\\
        \hline
        SEC& $\sigma^{eff} \geq 0 $ and $\sigma^{eff} +3p^{eff} \geq 0$ & $\mathcal{R}\leq \frac{(f_1-f_3-\theta-\theta\varphi)}{\varphi}$\\
          & & and $ (f_1-f_3-\theta-4\theta\varphi+\varphi \mathcal{R}) \geq 0$ \\
              \hline
              \end{tabular}
\end{table}

\section{Discussion}

Because of an additional higher-order curvature terms, modified gravitational theories are thought to be the most intriguing and promising way to study the present cosmic expansion. Here, with the geometric restriction of $P\left(GR_{4}\right)$ $\mathrm{GRW}$ spacetime, $f(\mathcal{R})$ gravity models are investigated and we find $p$ and $\sigma$ are not constants. Hence, we may assert that the $P\left(GR_{4}\right)$ $\mathrm{GRW}$ spacetime is consistent with the universe as it exists right now. Also, for the condition $f(\mathcal{R})=\mathcal{R}$, the spacetime reveals dust matter era, radiation era and dark energy epoch under certain restrictions. \par

The initial $f(\mathcal{R})$-model, $f(\mathcal{R}) = \mathcal{R} +\alpha \mathcal{R}^{2}$, ($\alpha>0$) proposed by Starobinsky \cite{aas} was aimed at explaining cosmic inflation as a pure gravitational effect without the use of dark energy. Carroll et al. \cite{car} presented the model $f(\mathcal{R}) = \mathcal{R}-\frac{\mu^{4}}{\mathcal{R}}$, ($\mu > 0$) to explain late-time acceleration as a scalar field. Despite their limitations, these models were still able to popularize $f(\mathcal{R})$-models in general. In \cite{LDMS21}, by choosing the model $f(\mathcal{R}) = \mathcal{R}-\alpha (1-e ^{-\frac{\mathcal{R}}{\alpha}})$ the authors have shown that DEC, NEC and WEC have been satisfied, whereas SEC violated. Here, our findings have been assessed both analytically and graphically. Our formulation was constructed using the analytical technique, and one cosmological model, $f\left(\mathcal{R}\right)=\mathcal{R}-\mu \mathcal{R}_{c}\tanh (\frac{\mathcal{R}}{\mathcal{R}_{c}})$, were evaluated for stability. For this model, we find that NEC, WEC, DEC, SEC and TEC are satisfied. Also, this finding yields the late-time acceleration of the Cosmos\cite{LDMS21} and each result aligns with the $\Lambda$CDM model \cite{AAAABBBBBBB20}.\par

In \cite{cap1}, Capozziello et al. deduced that a GRW spacetime of dimension $n$ with $\nabla_{k}C^{k}_{lij} =0$ reveals a perfect fluid type EMT for any $f(\mathcal{R})$ gravity model. Hence, from Theorem 2, we state
the subsequent:\par
A $P\left(GR_{4}\right)$ $\mathrm{GRW}$ spacetime represents a perfect fluid type EMT for any model of $f(\mathcal{R})$ gravity.\par

A collective study of validation of energy conditions are investigated and the result is mentioned in Table 1. It is seen that the presence of exotic matter is not required for our case.

\section{Declarations}
\subsection{Funding }
NA.
\subsection{Code availability}
NA.
\subsection{Availability of data}
NA.
\subsection{Conflicts of interest}
The authors have no conflicts to disclose.

\section*{Acknowledgment}
We would like to thank the referee and the Editor for reviewing the paper carefully and their valuable comments to improve the quality of the paper.\par


\begin{thebibliography}{00}	

\bibitem{Neil} B. O'Neil, Semi-Riemannian Geometry with Applications to the Relativity, Academic Press, New York-London, ISBN: 0125267401, 1983.
\bibitem{alias1}L. Alias, A. Romero, M. Sanchez,  Uniqueness of complete spacelike hypersurfaces of constant mean curvature in generalized Robertson-Walker spacetimes, Gen. Relativity Gravitation  27 (1995) 71-84.
\bibitem{bychen} B.Y. Chen,  Pseudo-Riemannian Geometry, $\delta$-invariants and Applications, World Scientific, 2011.
\bibitem{C14}B.Y. Chen, A simple characterization of generalized Robertson–Walker spacetimes, Gen. Relativity Gravitation 46 (2014) 1833.
\bibitem{MM16}C.A. Mantica, L.G. Molinari,  On the Weyl and Ricci tensors of Generalized Robertson-Walker space-times, J. Math. Phys. 57 (2016) 102502.
\bibitem{survey} C.A. Mantica, L.G. Molinari,  Generalized Robertson Walker spacetimes-A survey, Int. J. Geom. Methods Mod. Phys. 14 (2017) 1730001.
\bibitem{S98}M. S\'anchez,  On the geometry of generalized Robertson-Walker spacetimes: geodesics, Gen. Relativity Gravitation 30 (1998) 915-932.
\bibitem{HE73}S.W. Hawking, G.F.R. Ellis, The Large Scale Structure of Space-Time, Cambridge University Press, London, 1973.
\bibitem{chav} P.H. Chavanis,  Cosmology with a stiff matter era, Phys. Rev. D 92 (2015) 103004.
\bibitem{E49}L.P. Eisenhart,  Riemannian Geometry, Princeton University Press, ISBN: 9780691023533, 1949.
\bibitem{hj}H.J. Schmidt, Fourth order gravity: equations, history, and applications to cosmology, Int. J. Geom. Methods Mod. Phys. 04 (2007) 209.
\bibitem{P52}E.M. Patterson, Some theorems on Ricci-recurrent spaces, J. London Math. Soc. 27 (1952) 287–295.
\bibitem{DGK95}U.C. De, N. Guha, D. Kamilya,  On generalized Ricci-recurrent manifolds, Tensor (N.S.) 56 (1995) 312–317.
\bibitem{mdd}S. Mallick, A. De, U. C. De, On generalized Ricci recurrent manifolds with applications to relativity, Proc. Nat. Acad. Sci. India Sect. A 83 (2013) 143–152.
\bibitem{ade1}A. De, T.-H. Loo, R. Solanki, P. K. Sahoo,  A conformally flat generalized Ricci
recurrent spacetime in $f(R)$-gravity, Phys. Scr. 96 (2021) 085001.
\bibitem{ade2} A. De, T.-H. Loo, R. Solanki, P.K. Sahoo, How a conformally flat (GR)4 impacts Gauss-Bonnet
gravity? Fortschritte der Physik 69 (2021) 2100088.
\bibitem{ManSuh} C.A. Mantica, Y.J. Suh, Pseudo-Z symmetric space-times, J. Math. Phys. 55 (2014) 042502.
\bibitem{ozen} F.O. Zengin,  m-Projectively flat spacetimes, Math. Reports 14 (2012) 363-370.
\bibitem{fku} F. Mofarreh, K. De, U.C. De, Characterizations of a spacetime admitting $\Psi$-conformal curvature tensor,  Filomat 37 (2023) 10265–10274.
\bibitem{kdeu1}K. De, U.C. De, Some geometric and physical properties of pseudo $\psi$-conharmonically symmetric manifolds, Quaestiones Mathematicae 46 (2022) 939-958.
\bibitem{gul2} S. Güler, S.A. Demirbağ, A study of generalized quasi-Einstein spacetimes with applications in general relativity, Internat. J. Theoret. Phys. 55 (2016) 548–562.
\bibitem{kdez} P. Zhao, U.C. De, B. Unal, K. De, Sufficient conditions for a pseudosymmetric spacetime to be a perfect fluid spacetime, Int. J. Geom. Methods Mod. Phys. 18 (2021) 2150217.
\bibitem{BBHL07} O. Bertolami, C.G. B\"ohmer, T. Harko, F.S.N. Lobo, Extra force in $f\left(\mathcal{R}\right)$ modified theories of gravity, Phys. Rev. D 75 (2007) 104016.
\bibitem{CNO18} S. Capozziello, S. Nojiri, S.D. Odintsov, The role of energy conditions in $f\left(\mathcal{R}\right)$ cosmology, Phys. Letter B 781 (2018) 99-106.
\bibitem{EMOS10}E. Elizalde, R. Myrzakulov, V.V. Obukhov, D. S\'aez-G\'omez, $\Lambda$CDM epoch reconstruction from $F\left(\mathcal{R},G\right)$ and modified Gauss–Bonnet gravities, Classical Quantum Gravity 27 (2010) 095007.
\bibitem{HLNO11} T. Harko, F.S.N. Lobo, S. Nojiri S.D. Odintsov,  $f\left(R,T\right)$-gravity, Phys. Rev. D 84 (2011) 024020.
\bibitem{LPC15} M.D. Laurentis, M. Paolella, S. Capozziello, Cosmological inflation in $f\left(\mathcal{R},G\right)$ gravity, Phys. Rev. D,  91 (2015) 083531.
\bibitem{RBB92} A.K. Raychaudhuri, S. Banerji, A. Banerjee,  General relativity, astrophysics, and cosmology, Springer-Verlag New York, Inc., 1992.
\bibitem{DS99}K.L. Duggal, R. Sharma, Symmetries of Spacetimes and Riemannian Manifolds, Springer New York, NY, 1999.
\bibitem{tw} C. Barcelo, M. Visser, Twilight for the energy conditions?, Int. J. Mod. Phys. D
11 (2002) 1553.
\bibitem{hab} H.A. Buchdahl,  Non-linear Lagrangians and cosmological theory, Mon. Not. Roy. Astron. Soc. 150 (1970).
\bibitem{aas}A. A. Starobinsky, A new type of isotropic cosmological models without singularity, Phys. Letter B 91 (1980) 99-102.
\bibitem{har}T. Harko, F. S. N. Lobo, $f(R, L_{m})$-gravity, Eur. Phys. J. C 70 (2010) 373–379.
\bibitem {kat} N. Katirci,  M. Kavuk, $ f(R,T_{\mu\nu}T^{\mu\nu})$ gravity and Cardassian-like expansion as one of its consequences, Eur. Phys. J. Plus 129 (2014) 163.
\bibitem{cap}S. Capozziello, V. F. Cardone, V. Salzano,  Cosmography of $f(R)$ gravity, Phys. Rev. D 78 (2008) 063504.
\bibitem{cap2} S. Capozziello, R. D’Agostino, O. Luongo, Extended Gravity Cosmography, Int.
J. Mod. Phys. D (2019) doi:10.1142/S0218271819300167
\bibitem{ade}A. De, T.-H. Loo, S. Arora and P.K. Sahoo,  Energy condition for a $(W$RS$)_4$ spacetime in $f(R)$-gravity, Eur. Phys. J. Plus, 136 (2021) 218.

\bibitem{kde} K. De and U.C. De,  Investigations on solitons in $f(\mathcal{R})$-gravity,  Eur. Phys. J. Plus 137 (2022) 180.
\bibitem{kde1} U.C. De, K. De, F.O. Zengin, S.A. Demirbag,  Characterizations of a spacetime of quasi-constant sectional curvature and $\mathcal{F}(\mathcal{R})$-gravity, Fortschritte der Physik 71 (2023) 2200201.
\bibitem{shi} S. Tsujikawa,  Observational signatures of $f(R)$ dark energy models that satisfy cosmological and local gravity constraints, Phys. Rev. D, 77 (2008) 023507.
\bibitem{am}L. Amendola, S. Tsujikawa, Phantom crossing, equation-of-state singularities, and local gravity constraints in $f(R)$ models, Phys. Letter B 660 (2008) 125-132.
\bibitem{dol}A.D. Dolgov, M. Kawasaki, Can modified gravity explain accelerated cosmic expansion?,
Physics Letters B, 573 (2003) 1-4.
\bibitem{am1} L. Amendola, R. Gannouji, D. Polarski, S. Tsujikawa, Conditions for the cosmological viability of $f(R)$ dark energy models, Phys. Rev. D 75 (2007) 083504.
\bibitem{ys}Y.S. Song, W. Hu, I. Sawicki, Large scale structure of $f(R)$ gravity, Phys. Rev. D 75 (2007) 044004.
\bibitem{st}D.G. Boulware, S. Deser, String-Generated Gravity Models, Phys. Rev. Lett. 55 (1985) 2656.
\bibitem{Mantica3} C.A. Mantica, L.G. Molinari,  Weyl compatible tensors, Int. J. Geom. Methods Mod. Phys.  11 (2014) 1450070.
\bibitem{Bert} E. Bertschinger, A.J.S. Hamilton,  Lagrangian evolution of the Weyl tensor, Astroph.J.  435 (1994) 1-7.
\bibitem{S05}M. S\'anchez,  On the geometry of static spacetimes, Nonlinear Analysis: Theory, Methods \& Applications 63 (2005) 455-463.	

\bibitem{SKMHH09}H. Stephani, D. Kramer, M. Mac-Callum, C. Hoenselaers, E. Herlt,  Exact Solutions of Einstein’s Field Equations, Cambridge University Press, Cambridge, 2009.

\bibitem{Sotiriou} T.P. Sotiriou, V. Faraoni,  $f(R)$ theories of gravity, Modern Physics, 82 (2010) 451-497.

\bibitem{LDMS21}T.-H. Loo, A. De, S. Mandal, P.K. Sahoo, How a projectively flat geometry regulates F(R)-gravity theory? Phys. Scr. 96 (2021) 125034.

\bibitem{AAAABBBBBBB20}N. Aghanim, Y. Akrami, M. Ashdown, J. Aumont, C. Baccigalupi, M. Ballardini, A.J. Banday, R.B. Barreiro, N. Bartolo, S. Basak, R. Battye, Planck 2018
results-VI. Cosmological parameters, Astron. Astrophys. 641 (2020) A6.

\bibitem{car}S.M. Carroll, V. Duvvuri, M. Trodden, M. S. Turner,  Is cosmic speed-up due to new gravitational physics?, Phys. Rev. D 70 (2004) 043528.
\bibitem{cap1}S. Capozziello, C.A. Mantica, L.G. Molinari,  Cosmological perfect fluid in $f(R)$ gravity, Int. J. Geom. Methods Mod. Phys. 16 (2019) 1950008.
\end{thebibliography}
\end{document}